# Spin-ice behavior of three-dimensional inverse opal-like magnetic structures: micromagnetic simulations

I. S. Dubitskiy[1,2], A. V. Syromyatnikov[1,2], N. A. Grigoryeva[1], A. A. Mistonov[1,2], and S. V. Grigoriev[1,2]

[1]*Faculty of Physics, Saint-Petersburg State University, 198504 Saint Petersburg, Russia*
[2]*Petersburg Nuclear Physics Institute, Gatchina, 188300 Saint Petersburg, Russia*

We perform micromagnetic simulations of the magnetization distribution in inverse opal-like structures (IOLS) made from ferromagnetic materials (nickel and cobalt). It is shown that the unit cell of these complex structures, whose characteristic length is approximately 700 nm, can be divided into a set of structural elements some of which behave like Ising-like objects. A spin-ice behavior of IOLS is observed in a broad range of external magnetic fields. Numerical results describe successfully the experimental hysteresis curves of the magnetization in Ni- and Co-based IOLS. We conclude that ferromagnetic IOLS can be considered as the first realization of three-dimensional artificial spin ice.

## I. INTRODUCTION

A focus of interest in the nanomagnetism has shifted in the last decade from the standalone nanoobjects such as cubes [1], disks [2], pyramids [3, 4], hemispheres [5], octahedrons [6], and wires [7] to the ordered arrays of nanoobjects [8, 9] and porous ordered materials [10-12]. It has happened due to rapid progress in the lithographic [13] and self-assembling technologies [14]. One-dimensional (1D) and two-dimensional (2D) nanostructures such as antidot arrays are widely discussed now in magnonics [15-19]. A spin-ice behavior has been realized recently in 2D arrays of magnetic islands by a subtle tuning of system geometry [20, 21]. The latter finding opens a new perspective in discussion related to frustrated magnetic systems.

Possessing a topological phase with the fractional elementary excitations (magnetic monopoles) and macroscopic entropy in the low-temperature limit, spin ices have been intensively studied in recent years [22, 23, 24]. Pyrochlore materials such as $Dy_2Ti_2O_7$ and $Ho_2Ti_2O_7$ exhibit spin-ice properties caused by two reasons [22]. Firstly, the strong single-ion easy-axis anisotropy forces magnetic moments sitting at vertexes of corner-sharing tetrahedra to be directed to the center of one of the two tetrahedra. Secondly, the interaction between these moments is ferromagnetic. As a result, a frustration arises since ferromagnetic couplings of different moments compete with each other. Such systems are governed at small temperature by the so-called spin-ice rule (proposed initially for the water ice [25]). This rule states that a configuration with the minimal energy is achieved when two magnetic moments are directed into each tetrahedron and two moments are directed out of it (the 2-in-2-out rule). As the number of such states is macroscopically large in the pyrochlore lattice, the ground state is highly degenerate.

Artificial spin ice systems were initially designed to mimic spin-ice pyrochlores [26]. They consist of ordered single-domain ferromagnetic nanoparticles. The Ising-like behavior of magnetic moments of nanoparticles is provided by the shape anisotropy. However it is not easy to achieve the ground state degeneracy caused by the competing interactions in these systems [21, 26].

It was proposed recently that three-dimensional (3D) inverse opal-like structure (IOLS) made from ferromagnetic material can be the first realization of 3D spin-ice system [27]. To interpret the neutron scattering data in the Co-based IOLS, it was suggested that its magnetic behavior is determined mostly by Ising-like magnetic moments situated at vertices of tetrahedra-shaped structural units.

The present paper is aimed to provide a more solid confirmation of the spin-ice behavior of IOLS by studying magnetization distribution using micromagnetic simulations [28]. It is well known that magnetic properties of IOLS depend significantly on their geometry [29, 30, 31, 32]. Then, we restrict ourselves to Co- and Ni-based IOLS described in Refs. [27, 33] whose unit cells have a characteristic length of approximately 700 nm. We find that IOLS can be considered as a three-dimensional network of ferromagnetic nanoobjects (quasicubes and quasitetrahedra) which connect to each other via relatively long and thin crosspieces ("legs"). Due to their shape anisotropy, these legs possess the magnetic moments which can point either "in" or "out" nanoobjects. We find below that 2-in-2-out states of the tetrahedra-shaped structural units are realized in a broad range of external magnetic fields that results in highly degenerate ground state of the whole system. The situation here resembles that in pyrochlore materials $Dy_2Ti_2O_7$ and $Ho_2Ti_2O_7$ discussed above. Our numerical results describe successfully experimental hysteresis loops measured in Ni- and Co-based IOLS.

The spin-ice model proposed in Ref. [27] predicts an existence of several critical magnetic fields which correspond to magnetization reversal in the Ising-like structural units. However we find no abrupt changes of magnetization in the SQUID data which can be attributed to these critical fields. We show below that the demagnetizing field related to the shape of the whole sample smoothes out these peculiarities.

It should be noted that micromagnetic simulations play a very important role now in discussion of 3D magnetic structures with characteristic length of the structural unit of several hundred nanometers. Powerful experimental techniques such as Lorentz Microscopy [34], Magnetic Force Microscopy (MFM) [20] and X-ray Magnetic Circular Dichroism (XMCD) [35] not only make it possible to study the integral properties of 1D and 2D structures but also to investigate the magnetic state of each unit cell. However things become different when one decides to study, e.g., 3D antidot arrays. The surface sensitive methods mentioned above fail to study the 3D magnetization distribution in such objects. The SQUID and Vibrating Sample Magnetometer measurements can be used for studying the average value of the magnetization only. The Small-Angle Neutron Scattering (SANS) can in principle gain detailed information about the local magnetization distribution in bulk materials [27, 33, 36]. Sometimes, however, the SANS resolution is insufficient for investigation of subtle details of structures built from objects with the characteristic lengths of several hundred nanometers. Due to the lack of experimental techniques which can directly restore a three-dimensional magnetization distribution, the numerical calculations should be employed to gain insight into the magnetism of these complex structures and to interpret properly experimental data. The most straightforward and rigorous way of magnetic nanostructures modeling is solving the Landau–Lifshitz–Gilbert (LLG) equation [3, 6, 37, 38, 39].

The rest of the present paper is organized as follows. The sample fabrication and characterization methods are briefly discussed in Section II. Section III is devoted to the inverse opal geometry which is essential for understanding of the magnetic behavior of IOLS. Micromagnetic modeling technique is introduced in Section IV. Results of calculations and comparison with experimental data are presented in Section V. Summary and concluding remarks are given in Section VI.

## II. SAMPLE PREPARATION AND EXPERIMENTAL METHODS

The Ni- and Co-based IOLS have been fabricated by means of the templating technique [40]. Firstly, a colloidal crystal was formed from polystyrene microspheres with the mean diameter of about 500 nm by the vertical deposition technique [41]. The microspheres were arranged in the face-centered cubic (fcc) packing [42]. After that, the electrochemical deposition of the nickel or cobalt into the voids between microspheres was carried out in the three-electrode cell. Finally, the polystyrene microspheres were dissolved in the toluene.

The scanning electron microscope (SEM) images of the IOLS surface and the side face are presented in Fig. 1. The small black spots in the spherical grey voids are caused by deformation (or sintering) of the microspheres which form the colloidal crystal. The IOLS surface corresponds to (111) plane of the fcc structure. It was shown by Small-Angle X-ray Scattering (SAXS) that the IOLS has the same fcc symmetry as the template colloidal crystal and it has the period of about 700 nm [33, 41, 43, 44].

According to the wide angle Powder X-ray Diffraction experiments, the IOLS based on nickel consists of polycrystalline fcc nickel grains [33, 43] whereas the cobalt-based IOLS consists of polycrystalline hexagonal close-packed (hcp) cobalt grains [44].

Hysteresis curves presented below have been measured by the SQUID magnetometer Quantum Design MPMS-5S at Institute of Condensed Matter Physics, Braunschweig, Germany. The magnetic field ranging from -5 T to +5 T was applied along [111] axis of the IOLS, i.e., perpendicular to the film plane. The same configuration of the field was used in the micromagnetic modeling.

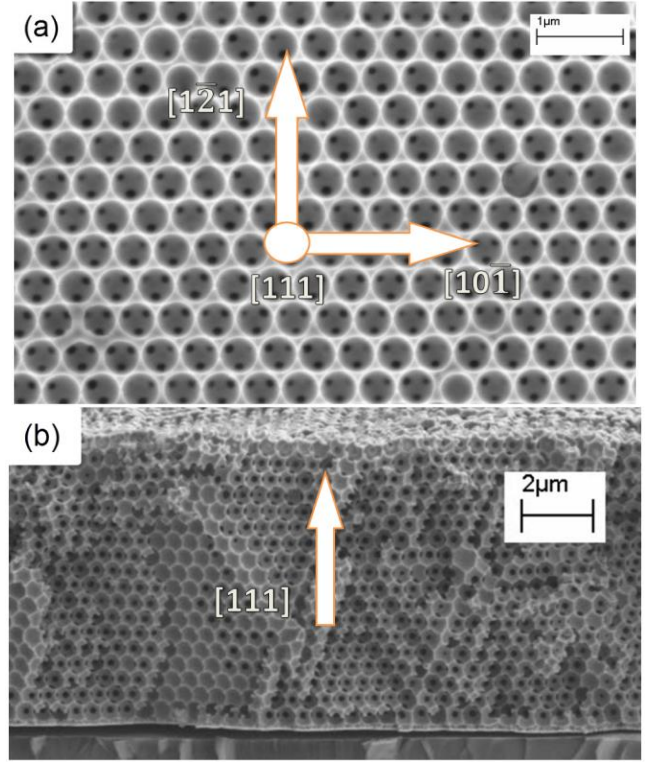

FIG. 1. SEM images of the Ni-based IOLS: (a) top view corresponding to (111) plane and (b) side view. Arrows indicate [1$\bar{2}$1], [10$\bar{1}$], and [111] crystallographic directions.

## III. INVERSE OPAL GEOMETRY

Though inverse opals possess the fcc symmetry, the IOLS unit cell looks very unusual for an ordinary fcc structure (see Fig. 2(a) and Fig. 2(b)). Two types of voids, octahedral-shaped and tetrahedral-shaped, constitute the unit cell of this fcc structure. Additionally, microspheres do not have just the point-like contacts between each other in real colloidal crystals. They are deformed (or sintered) a little bit [45]. As a result of this sintering, the octahedral and the tetrahedral voids are connected by thin elongated crosspieces. We denote them as "legs". All the legs are oriented along <111> crystallographic axes. Therefore, the inverse fcc structure can be considered as an ordered array composed of two different types of elements connected to each other by legs. Elements of the first type are formed by the octahedral voids. Following Ref. [27], we denote them as quasicubes. The elements of the second type are formed by the tetrahedral voids (we denote them as quasitetrahedra).

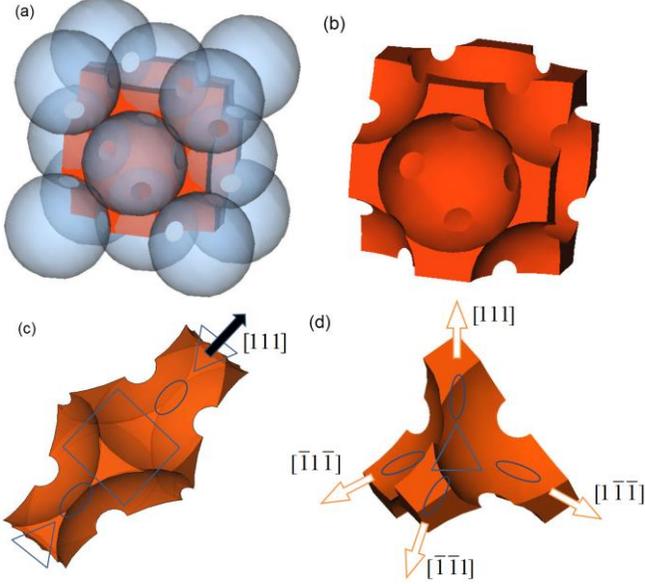

FIG. 2. The unit cell of the considered IOLS with (a) and without (b) the surrounding microspheres of the colloidal crystal (the sintering degree is 2%). The primitive cell of the IOLS (c) and the unit cell octant (d). Squares and triangles indicate the octahedral (quasicubes) and tetrahedral (quasitetrahedra) voids, respectively. Ellipses indicate "legs".

For example, the IOLS primitive unit cell (i.e., the cell which is built on the primitive fcc basis) consists of two differently oriented quasitetrahedra, one quasicube between them, and two legs connecting these three elements (see Fig. 2(c)). Bases of both quasitetrahedra lay in (111) plane but their heights are oriented along [111] and [$\bar{1}\bar{1}\bar{1}$] axes, respectively. It is shown below that the quasitetrahedra surrounded by four legs (Fig. 2d) play the key role in the process of the magnetization reversal in the unit cell of IOLS.

It should be noted that the shape of the IOLS unit cell depends strongly on the degree of the microspheres sintering which we define as $k-1$ expressed in percent, where $k = r'/r$, $r'$ is the sphere radius, and $r$ is a half of distance between the spheres centers. For example, Fig. 3 shows the unit cell of the IOLS built on the template of 10% sintering microspheres. It is seen that legs between quasicubes and quasitetrahedra become narrower and longer upon the sintering increasing (cf. Fig. 2(b), Fig. 2(c) and Fig. 3(a), Fig. 3(b)). One can roughly estimate a leg length as a diameter of circle formed by a sphere intersection. It is equal to 100 nm for the considered IOLS shown in fig. 2 which sintering degree is 2%. We should point out that legs are formed exclusively as a result of microspheres sintering. Hence, "ideal" inverse opals made from perfect microspheres could not be considered as spin ices in any case. It is shown in [46] that electropolishing can be used in order to effectively increase the sintering degree. The unit cell presented in FIG. 2(b) has been used in micromagnetic calculations.

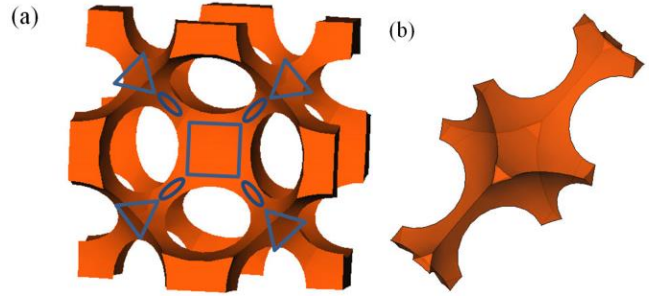

FIG. 3. Unit cell of the IOLS (a) and the primitive cell (b) at high sintering degree (10%). Squares and triangles indicate octahedral and tetrahedral voids, respectively. Ellipses indicate "legs".

## IV. MICROMAGNETIC MODELLING

We use Nmag micromagnetic simulation package [47] provided by University of Southampton. This software performs the time integration of the LLG equation using the so-called Finite Element Method/Boundary Element Method [48]. The hierarchical matrices are used for the boundary element matrix approximation [49]. The LLG equation can be written as follows:

$$\frac{\partial \mathbf{M}}{\partial t} = -\gamma \mathbf{M} \times \mathbf{H}_{\text{eff}} + \frac{\alpha}{M_S} \mathbf{M} \times \frac{\partial \mathbf{M}}{\partial t}, \qquad (1)$$

where **M** is the magnetization vector which depends on the space coordinates and time, $\gamma$ is the Gilbert gyromagnetic ratio, $\alpha$ is the phenomenological damping parameter, $\mathbf{H}_{eff}$ is the effective field. The effective field is defined as a negative variational derivative of the free magnetic energy with respect to the magnetization. The free magnetic energy is considered as a sum of the exchange energy, the demagnetization energy, and the Zeeman energy (the energy of the system in external magnetic field). Since we do not consider the system dynamics, we do not take the precession into account and put the damping parameter to be equal to unity.

We do not take into account a magnetocrystalline anisotropy since our samples are polycrystalline (see Section II). We use the following material parameters in our simulations: the exchange coupling constant $A = 10^{-11}$ J/m and the saturation magnetization $M_S = 4.82 \cdot 10^5$ A/m for the Ni-based IOLS, and $A = 3 \cdot 10^{-11}$ J/m, $M_S = 1.4 \cdot 10^6$ A/m for the IOLS based on cobalt [50, 51, 52]. The exchange length can be calculated from the equation $l_{ex} = \sqrt{2A/(\mu_0 M_S^2)}$ [50, 53]. It is equal to 8.3 nm and 4.9 nm for nickel and cobalt samples, respectively.

The linear size of the finite element is the key parameter that controls the accuracy of the micromagnetic simulation. One has to choose it smaller than the exchange length in order to obtain reasonable results [54]. However the linear size of the IOLS unit cell is of the order of 700 nm. The modeling of the magnetization distribution in such a big object is a very computationally challenging task even for a high performance computer cluster. As a result, it is impossible to calculate the magnetization distribution in a reasonable space of time in the whole sample or even in several unit cells. Then, we have to restrict ourselves to simulation of one unit cell only. It takes for about a month to calculate a hysteresis loop containing 41 points on a eight node SMP machine. The typical value of RAM used is 50 GB. It should be noted that periodical boundary conditions (PBC) or even the macrogeometry approach [55] can be used to take periodicity into account. Nevertheless, we think that it is not necessary to do at least in the first approximation due to the following reasons. Firstly, SANS data [27, 33] shows no magnetization periodicity near the coercivity point. Therefore, open boundary conditions (OBC) seem to be more appropriate in this region than PBC. This assumption is supported by the reasonably good coincidence between calculated and experimental curves at small fields (see Fig. 16 and Fig. 17). Secondly, preliminary calculations suggest that the difference between OBC and PBC is rather small in high field (several percent).

We have to take into account the demagnetizing field related to the sample shape in order to compare the experimental data with the results of the modeling. Our samples have the form of a thin film which demagnetizing tensor has the only nonzero *zz*-component that is equal to 1 (*z* axis is perpendicular to the IOLS surface and it coincides with [111] crystallographic direction as is shown in Fig. 1). We have for the z-component of the demagnetizing field

$$H_z^{dem} = -M_z \frac{V_{IOLS}}{V_{film}}, \qquad (2)$$

where $M_z$ is the z-component of the average magnetization per unit volume, $V_{IOLS}$ is the volume of the magnetic material in the IOLS, and $V_{film}$ is the volume of the sample. If spheres of the initial opal template only touch each other, we obtain for the ratio $V_{IOLS}/V_{film} = 1 - \pi/(3\sqrt{2}) \approx 0.26$. Taking sintering into account, one derives the following analytical expression:

$$\frac{V_{IOLS}}{V_{film}} = 1 - \frac{\pi}{\sqrt{2}}\left(\frac{k^3}{3} - 3k(k-1)^2 + (k-1)^2\right), \quad (3)$$

where $k = r'/r$.

It is difficult to extract $k$ from experimental data. As it is noted above, the resolution of the modern small angle scattering setups is limited. We have used SEM images to estimate $k$ and have found that it is equal approximately to 1.02. Then, the degree of the sintering is equal to 2% (see Fig. 2). Eq. (3) gives in this case $V_{IOLS}/V_{film} \approx 0.22$.

The demagnetizing magnetic field is parallel to the external one $H_{ext}$ when the latter is applied along [111] axis. We consider this simple case below as it is of particular interest for the ice-rule concept verification. As soon as [111] crystallographic direction corresponds to the unit cell diagonal, a 2-in-2-out state turns out to be unfavorable for the Zeeman energy because not all angles between magnetic moments and the applied field are smaller than 90 degrees (see Fig. 2(d)). Hence, it is expected that the ice rule is violated at large field. But it could be restored during the field decrease. The intrinsic field $H_{in}$ acting on the unit cell and the external field are related to each other as follows [56]:

$$H_{ext} = H_{in} + 0.22 M_z. \qquad (4)$$

For our simulations by the Finite Element Method/Boundary Element Method, we divide the unit cell volume into a number of finite elements with a maximal linear size equal to 5 nm. Slight variations of the finite elements size and of the mesh granularity do not affect results of simulation.

## V. RESULTS AND DISCUSSION

In general, the magnetization distribution in the unit cell is rather complicated. Fig. 4(a) displays the remanence state in the Ni-based IOLS unit cell and in its quasicube (Fig. 4(b)). Note that we use the coordinate system based on the fcc structure basis in the unit cell consideration. A vortex state is set in the quasicube with a bent vortex core. The vortex state is observed in the whole range of fields between two merging points of the hysteresis branches. Then, magnetization distribution is highly nonuniform in quasicube and its contribution is small to the net magnetization.

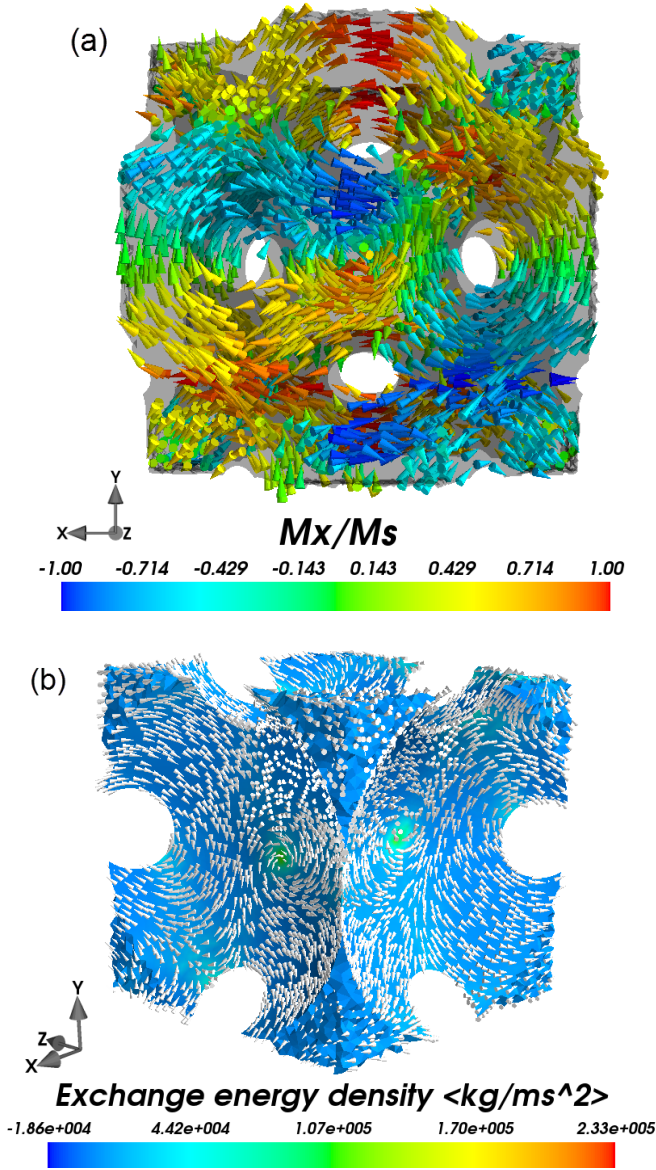

FIG. 4.    (Color online.) The magnetization distribution in the unit cell of the Ni-based IOLS (a) (color indicates the value of x-component of the normalized magnetization) and in its quasicube (b) (color displays the exchange energy density) in the remanence state.

At the same time, results of the micromagnetic modelling suggest that the magnetization distribution in IOLS legs is uniform in a wide range of fields. The magnetization along four legs of the typical quasitetrahedron is shown in Fig. Fig. 5 and Fig. Fig. 6 as a function of applied field for Ni- and Co-based IOLS, respectively. For definiteness, we consider the decreasing branch of the hysteresis curve. Signs of the magnetization projections are taken according to directions shown in Fig. 2(d).

One can see that magnetic moments of legs behave as Ising-like moments which point either in or out of quasitetrahedra. At large magnetic field, all quasitetrahedra are in 3-in-1-out or 3-out-1-in states. Hence, the 2-in-2-out ice-rule configuration is destroyed by strong external magnetic field. Conversely, in small fields one of 2-in-2-out configuration arises in quasitetrahedra.

However Fig. 5 and Fig. 6 show some peculiarities. Firstly, quasitetrahedra do not switch immediately from 2-in-2-out to 3-out-1-in (or 3-in-1-out) states at negative external magnetic fields. Magnetization in the corresponding leg inclines towards field direction before the switching. An example of such a nonuniform state is shown in Fig. 7(a). These states take place both in Ni- and Co-based IOLS. Evidently, such states occur due to insufficiently strong shape anisotropy of IOLS legs. Presumably, the shape anisotropy could be increased by increasing the sintering degree. However, it is out of the scope of the present paper to consider in more detail the role of the sintering. Secondly, magnetic moments in legs which are not parallel to the external magnetic field may reverse through a vortex state in IOLS based on Co (see Fig. 7(b)). Magnetization of legs in this state is close to zero (see the field range around 100 mT for $[\bar{1}1\bar{1}]$ leg in Fig. 6). However, such states exist in rather narrow ranges of fields which are less than 20 mT. Such vortex states are not observed in Ni-based IOLS that can be explained by larger exchange length in Ni (8.3 nm) than in Co (4.9 nm).

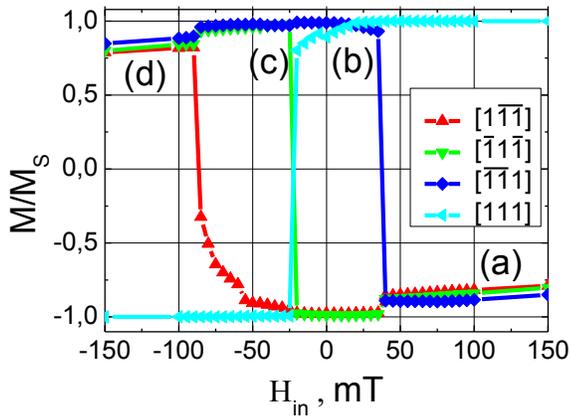

FIG. 5. (Color online.) The magnetization along the quasitetrahedron legs for the Ni-based IOLS. Legs are denoted by the corresponding <111>-type axes. Letters (a), (b), (c) and (d) mark quasitetrahedron magnetic states shown in Fig. 14.

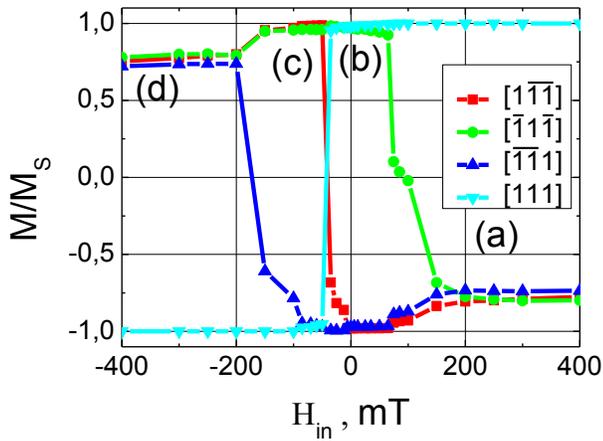

FIG. 6. (Color online.) The magnetization along the quasitetrahedron legs for the Co-based IOLS. Legs are denoted by the corresponding <111>-type axes. Letters (a), (b), (c) and (d) mark quasitetrahedron magnetic states shown in Fig. 15.

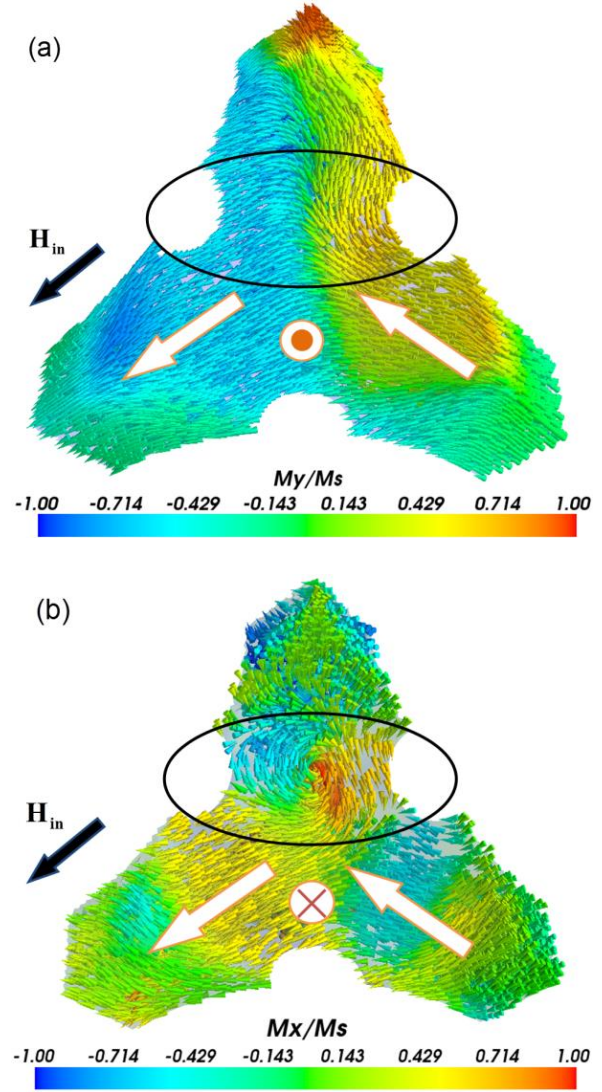

FIG. 7. (Color online.) (a) The nonuniform magnetization of the leg (indicated by ellipse) of the nickel quasitetrahedron at -85 mT on the decreasing branch of the hysteresis curve. (b) The vortex in the leg of the cobalt quasitetrahedron at 100 mT on the decreasing branch of the hysteresis curve. White arrows display directions of the average magnetic moments in legs. The direction of the magnetic moment in the fourth leg is shown either by the cross or by the dot.

In order to gain insight into the origin of 2-in-2-out states, it is necessary to analyze the behavior of the exchange and the demagnetization energies during the reversal processes in legs. We consider the total exchange and demagnetization energies of quasitetrahedra as functions of applied magnetic field in the decreasing branch of hysteresis curves. Hereinafter letters (a), (b), (c) and (d) mark quasitetrahedron magnetic states shown in Fig. 14 and Fig. 15

The exchange energy drops both in Co- and Ni-based quasitetrahedra during the switching from 3-in-1-out (or 3-out-1-in) to 2-in-2-out states (see regions around 40 mT and 80 mT in Fig. 8 and Fig. 11, respectively). In Co-based quasitetrahedron, additional substantial increase of the exchange energy just before switching takes place due to formation of the vortex state in the leg (see Fig. 7(b)). Different realizations of 2-in-2-out states at the same field value possess the same energy (see regions (b) and (c) in Fig. 8 and Fig. 11). Hence, the ground state degeneracy arises The demagnetization energy also drops in 2-in-2-out state (see Fig. 9 and Fig. 12). However due to non-local nature of the demagnetizing field, the whole unit cell contributes to quasitetrahedron demagnetization energy. Finally the magnetic charge $\int_{V_{qtetr}} \mathrm{div}(\mathbf{M}) dV$ of quasitetrahedra is more than an order of magnitude larger in 3-out-1-in (or 3-in-1-out) state than in a 2-in-2-out configuration (see fig. 10 and fig. 13). It is well known that the formation of the magnetic charge increases the demagnetization energy [24, 57].

Overall, one concludes that 2-in-2-out state minimizes both exchange and demagnetization energies. Thus, no competing arises between them. At the same time the IOLS ground state remains degenerate due to large number of 2-in-2-out configurations having the same energy.

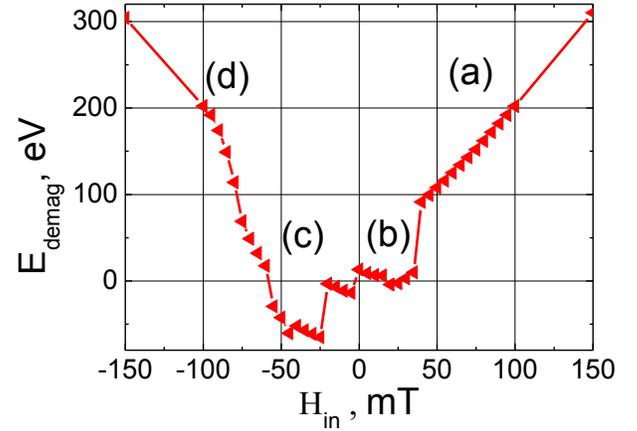

FIG. 9.   (Color online.) Total demagnetization energy of a quasitetrahedron in Ni-based IOLS

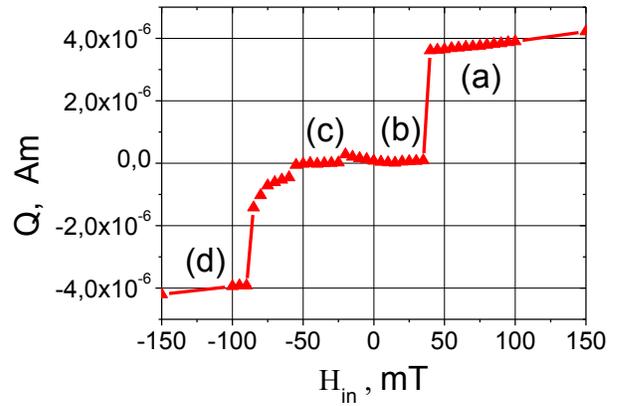

FIG. 10.   (Color online.) Total magnetic charge of a quasitetrahedron in Ni-based IOLS

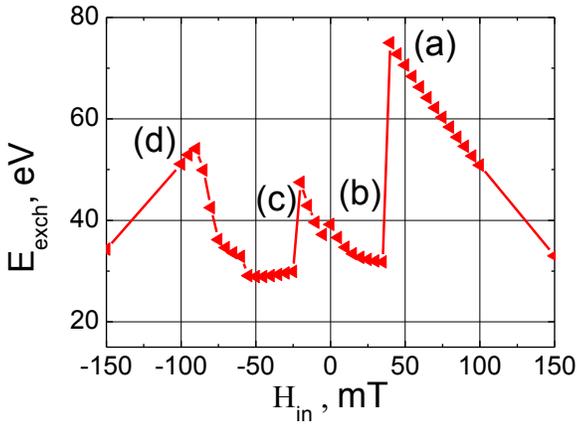

FIG. 8.   (Color online.) Total exchange energy of a quasitetrahedron in Ni-based IOLS.

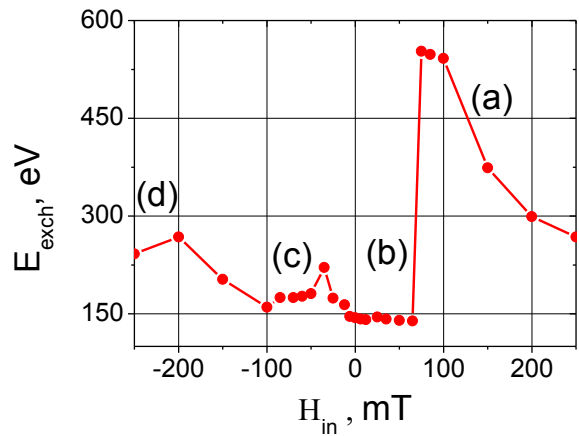

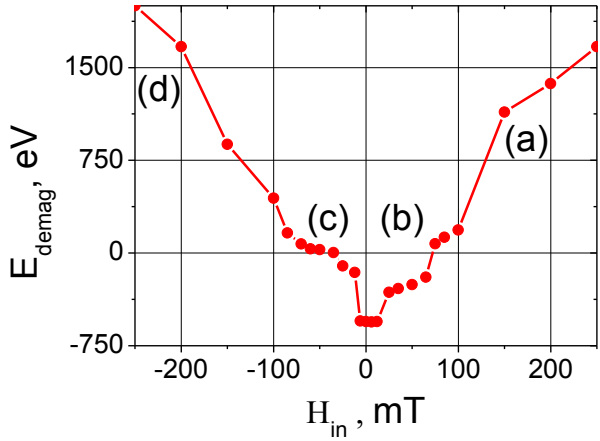

FIG. 11. (Color online.) Total exchange energy of a quasitetrahedron in Co-based IOLS

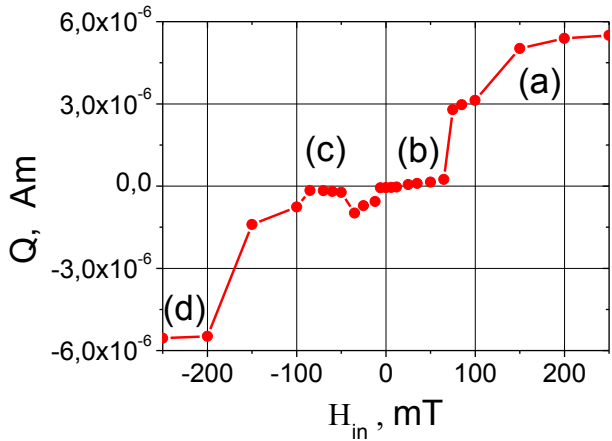

FIG. 12. (Color online.) Total demagnetization energy of a quasitetrahedron in Co-based IOLS

FIG. 13. (Color online.) Total magnetic charge of a quasitetrahedron in Co-based IOLS

Let us turn now to the analysis of the whole unit cell hysteresis loop.

Based on the micromagnetic modeling results, we can conclude that abrupt irreversible changes of the magnetization distribution in the unit cell are caused by reversals of legs magnetic moments. Calculated hysteresis loops and the magnetization distribution in the typical quasitetrahedron of the Ni- and Co-based IOLS are shown in Fig. 14 and Fig. 15, respectively.

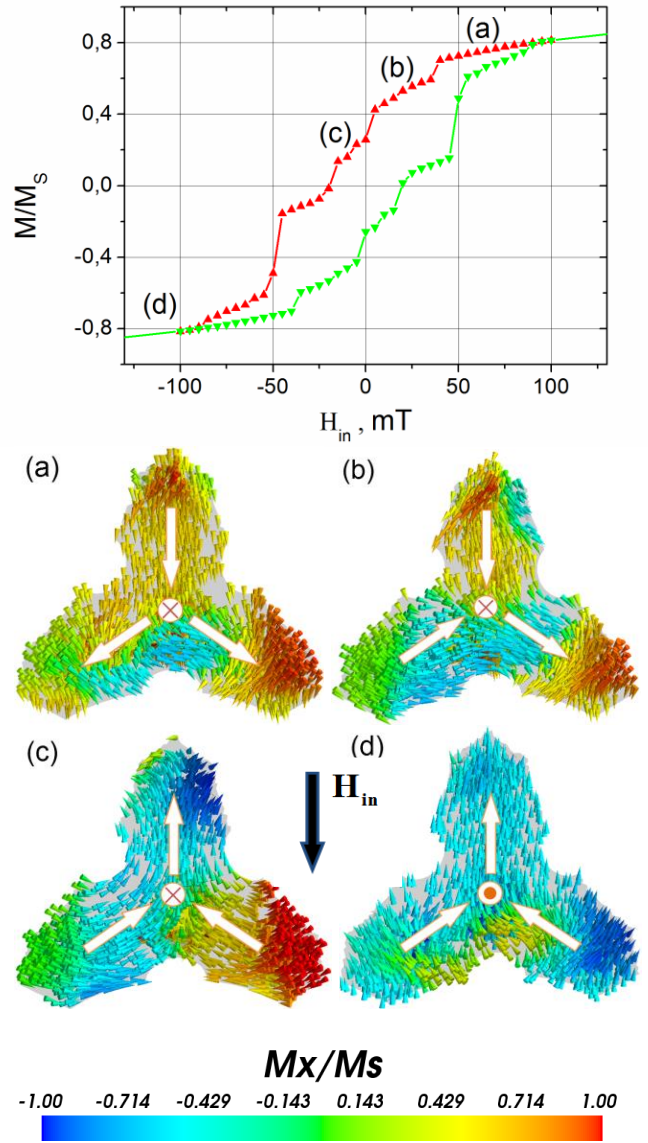

FIG. 14. (Color online.) The hysteresis loop of the Ni-based IOLS unit cell in the external field applied along [111] direction. For selected parts of the curve, the magnetization distributions in the typical quasitetrahedron are shown in panels (a)-(d). (a) 3-out-1-in state (half of quasitetrahedra are in 3-in-1-out state). (b) and (c) different realizations of the 2-in-2-out state. (d) 3-in-1-out state (half of quasitetrahedra are in 3-out-1-in state). Color indicates the value of x-component of the normalized magnetization. White arrows display directions of average magnetic moments in legs. The direction of the magnetic moment in the fourth leg is shown either by cross or by dot. The direction of the external field is shown by black vertical arrow.

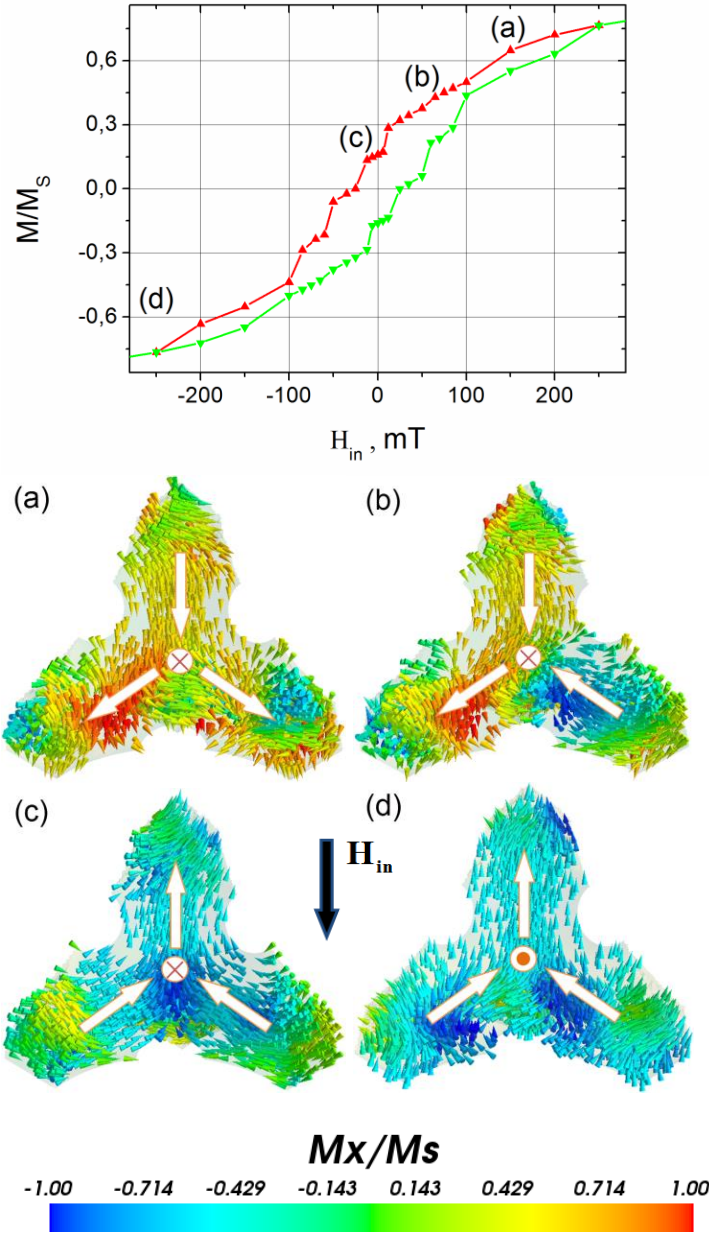

FIG. 15. (Color online.) The same as in Fig. 14 but for the unit cell of the Co-based IOLS.

Hysteresis loops are characterized by a number of magnetization jumps. At large external fields (larger than 300 mT and 100 mT for Co- and Ni-based IOLS, respectively), all quasitetrahedra are in 3-in-1-out or 3-out-1-in states (panels (a) and (d) in Fig. 14 and Fig. 15). The first jump occurs upon the field decreasing to 150 mT and 35 mT in Co- and Ni-based IOLS, correspondingly. A reorientation of magnetic moments in legs takes place at those fields that leads to 2-in-2-out states in six quasitetrahedra in the Co-based IOLS and in four quasitetrahedra in the Ni-based IOLS (typical 2-in-2-out states obeying the spin-ice rule are shown in Fig. 14(b) and Fig. 15(b)). All quasitetrahedra fall in 2-in-2-out configurations in the external field of 50 mT and -25 mT in IOLS based on Co and Ni, respectively. The difference between these field values stems mainly from different exchange length values and in Co and Ni.

Indeed, it is easier for a leg to switch the direction of its magnetic moment (via a vortex state) if the exchange length is sufficiently small. The particular geometry of legs and the particular value of the exchange length make possible (impossible) the formation of the intermediate vortex state in legs of Co- (Ni-) based IOLS (see Fig. 7). It should be noted that vortex state was observed in high field only during the transition from the 3-in-1-out (or 3-out-1-in) to 2-in-2-out state. Overall the advantage of 2-in-2-out configurations can be increased by subtle tuning the exchange length (i.e., by the proper choice of the material) and the legs geometry (by varying the sintering degree and the period of the structure).

Other jumps on hysteresis loops are related to magnetization reversals in legs which are parallel to the external field ([111] direction). Remarkably, when the magnetic moment in one leg reverses, other magnetic moments in legs of the same quasitetrahedron change their direction so that a 2-in-2-out configuration realizes (see, e.g., the transition from (b) to (c) in Fig. 5 and Fig. 6). Overall, we suggest that 2-in-2-out configurations are favorable for both IOLS.

Using Eq. (4), we can easily relate the internal field $H_{in}$ applied to the unit cell during micromagnetic simulation to the external field $H_{ext}$ applied to the sample in the experiment. One can directly compare the numerical and experimental results in this way.

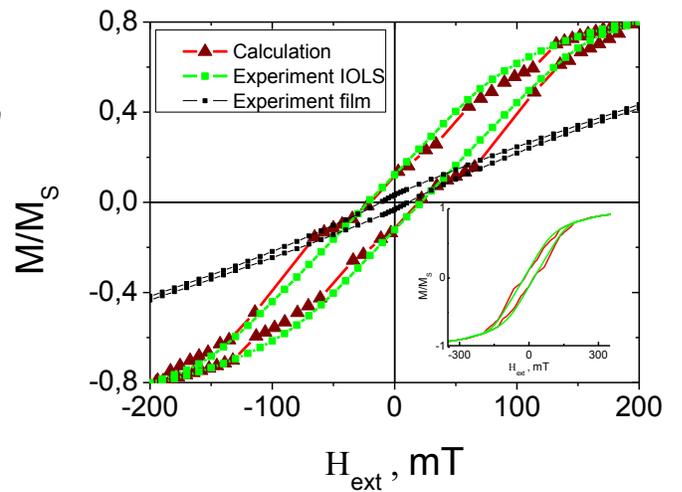

FIG. 16. (Color online.) The experimental and calculated hysteresis loops for the Ni-based IOLS. The external field is applied along [111] direction. Inset shows the hysteresis loops in a wider field range. Experimental data for Ni 3.2 μm film is also shown for comparison.

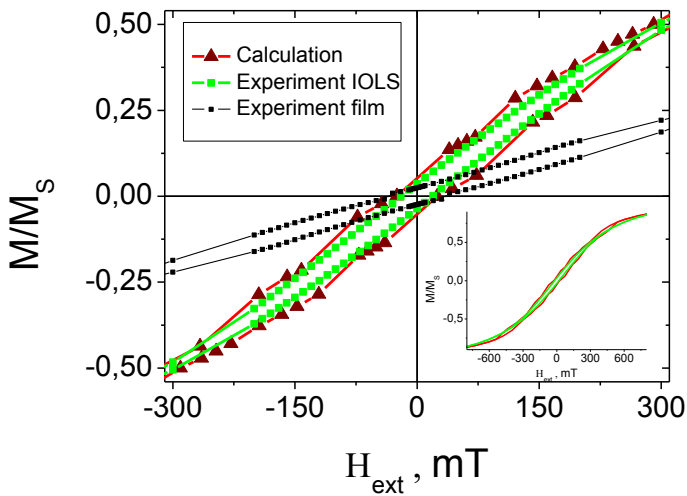

FIG. 17. (Color online.) The same as in Fig. 8 but for the Co-based IOLS and the Co film.

Fig. 16 and Fig. 17 display the calculated and experimental hysteresis loops for Ni- and Co-based IOLS, respectively. Experimental data for the continuous Co and Ni films are also presented for comparison. Comparing Fig. 16 and Fig. 17 with Fig. 14 and Fig. 15, one can see that demagnetizing field makes hysteresis loop smoother and it eliminates jumps almost completely. A significant difference should be pointed out also between experimental data obtained in continuous films and in IOLS. In general, this is connected with much larger perpendicular anisotropy of IOLS. This anisotropy in turn is caused by the shape anisotropy of IOLS legs. The experimental and calculated hysteresis loops coincide particularly well at small external fields (in the range of [-40 ÷ 40] mT and [-70 ÷ 70] mT for the Ni- and Co-based IOLS, correspondingly) when the sample magnetization is small and no long-range magnetic order exists in IOLS according to SANS data [27].

## VI. SUMMARY AND CONCLUSION

To conclude, the micromagnetic modeling of the magnetization distribution is performed in inverse opals on the base of cobalt and nickel ferromagnets. The external magnetic field is applied along [111] crystallographic axis of the IOLS corresponding to the cube diagonal of the fcc structure. We show that the inverse opal can be considered as a three-dimensional network of ferromagnetic nanoobjects (quasicubes and quasitetrahedra) which connect to each other via relatively long and thin crosspieces ("legs"). We find that due to their shape anisotropy these legs possess magnetic moments which can point either "in" or "out" nanoobjects in a wide range of fields. Therefore, magnetic moments of legs can be considered as Ising-like objects. We obtain that 2-in-2-out states of quasitetrahedra minimize both the exchange and the demagnetization energies of IOLS in a broad range of applied fields leading to a macroscopic degeneracy of the ground state. Thus, we observe two necessary attributes of an artificial spin ice: presence of the Ising-like moments and the spin-ice rule fulfillment.

The 2-in-2-out configurations turn into 3-in-1-out (or 3-out-1-in) states at large magnetic fields which values are close to those at which hysteresis branches merge. Similar transitions take place in pyrochlore spin ices [58]. We demonstrate also that 2-in-2-out states are more favorable for the Co-based IOLS than for the Ni-based IOLS. This is related to the smaller exchange length in cobalt that results in the ability of legs to switch the direction of their magnetization via vortex intermediate states. Then, we conclude that one can increase the quality of 3D artificial spin ice on the base of inverse opals by tuning the legs geometry and the structure period as well as by proper choice of the material (the exchange length).

We compare our numerical results with SQUID data and find a good agreement between them.

## ACKNOWLEDGMENTS

The calculations were carried out using the computer resources provided by the Resource Center "Computer Center of SPbU" (http://cc.spbu.ru). We thank N. Sapoletova, K. Napolsky, and A. Eliseev for the sample fabrication. We are grateful to I. Shishkin and D. Menzel for SQUID measurements and discussions. We acknowledge the assistance of Interdisciplinary Resource Center for Nanotechnology for the SEM investigations.

This work was supported in part by the RFBR projects No. 14-22-01113. I.S.D. and A.V.S. acknowledge Russian Scientific Fund Grant No. 14-22-00281 for financial support.